\documentclass[8.5pt,twoside,twocolumn]{article}
\usepackage[super,sort&compress,comma]{natbib} 
\usepackage{mhchem}
\usepackage{times,mathptmx}
\usepackage{sectsty}
\usepackage{balance} 

\usepackage{graphicx} 
\usepackage{lastpage}
\usepackage[format=plain,justification=raggedright,singlelinecheck=false,font=small,labelfont=bf,labelsep=space]{caption} 

\begin{document}

\makeatletter 
\def\subsubsection{\@startsection{subsubsection}{3}{10pt}{-1.25ex plus -1ex minus -.1ex}{0ex plus 0ex}{\normalsize\bf}} 
\def\paragraph{\@startsection{paragraph}{4}{10pt}{-1.25ex plus -1ex minus -.1ex}{0ex plus 0ex}{\normalsize\textit}} 
\renewcommand\@biblabel[1]{#1}            
\renewcommand\@makefntext[1]%
{\noindent\makebox[0pt][r]{\@thefnmark\,}#1}
\makeatother 
\renewcommand{\figurename}{\small{Fig.}~}
\sectionfont{\large}
\subsectionfont{\normalsize} 

\twocolumn[
  \begin{@twocolumnfalse}
\noindent\LARGE{\textbf{Velocity map imaging of a slow beam of ammonia molecules inside a quadrupole guide}}
\vspace{0.6cm}

\noindent\large{\textbf{Marina Quintero-P\'{e}rez\textit{$^\ddag$},
Paul Jansen\textit{$^\ddag$} and
Hendrick L. Bethlem}}\vspace{0.5cm}

\vspace{0.6cm}

\noindent \normalsize{Velocity map imaging inside an electrostatic quadrupole guide is demonstrated. By switching the voltages that are applied to the rods, the quadrupole can be used for guiding Stark decelerated molecules and for extracting the ions. The extraction field is homogeneous along the axis of the quadrupole while it defocuses the ions in the direction perpendicular to both the axis of the quadrupole and the axis of the ion optics. To compensate for this astigmatism, a series of planar electrodes with horizontal and vertical slits is used. A velocity resolution of 35\,m/s is obtained. It is shown that signal due to thermal background can be eliminated, resulting in the detection of slow molecules with an increased signal-to-noise ratio. As an illustration of the resolving power, we have used the velocity map imaging system to characterize the phase-space distribution of a Stark decelerated ammonia beam.}
\vspace{0.5cm}
\end{@twocolumnfalse}]

\section{Introduction}
\footnotetext{\textit{Institute for Lasers, Life and Biophotonics, VU University Amsterdam,
de Boelelaan 1081, 1081 HV Amsterdam, The Netherlands.}}
\footnotetext{\ddag~These authors contributed equally to this work}

Techniques to decelerate and cool molecules increase the interaction time in spectroscopic experiments, and promise to significantly enhance the precision of a number of experiments aimed at testing fundamental laws of physics\cite{Hudson:Nature2011,Tarbutt:FarDis2009,DeMille:PRL2008,Bethlem:EPJST2008}. 
As a proof of principle, high resolution microwave spectroscopy was carried out on Stark decelerated molecular beams of ND$_3$\cite{vanVeldhoven:EPJD2004} and OH\cite{Hudson:PRL2006}. In these experiments, an interaction time on the order of a millisecond was achieved. In order to increase the interaction time further, a molecular fountain is currently being constructed in our laboratory\cite{Bethlem:EPJST2008}. Unfortunately, due to the limited efficiency of cooling techniques, the increased interaction time comes at the expense of a decreased signal-to-noise ratio. Techniques such as resonance enhanced multi-photon ionization (REMPI) allow detection of very small quantities of molecules. However, if the signal of the decelerated molecules becomes too small, it will be obscured by the signal of the undecelerated part of the beam.

Here, we use velocity map imaging ({\sc{vmi}}) to discriminate the signal of slow molecules from thermal background molecules. {\sc{vmi}} is a technique that uses ion lenses to focus laser-produced ions onto a position dependent ion detector\cite{Eppink:RevSciInstrum1997}. By setting the voltages of the ion lenses correctly, molecules that have the same initial velocity but a different initial position are focused at the same position on the detector. In this way, all velocity information of the particles involved in the experiment is contained in a single image. {\sc{vmi}} or, more generally, ion imaging techniques have been used extensively for measuring the velocity of product molecules or atoms following a chemical reaction or the photodissociation of a parent molecule\cite{Chandler:JCP1987,Eppink:RevSciInstrum1997,Whitaker:Book}. State-of-the-art velocity map imaging systems achieve a velocity resolution below 10\,m/s\cite{Lipciuc:PCCP2006}. 

The ion optics used in a conventional {\sc{vmi}} setup consist of three or more electrodes with circular apertures. In this paper, we perform velocity map imaging inside an electrostatic quadrupole guide that is mounted directly behind a Stark decelerator. 
The main advantage of this configuration is that it allows for detection inside the guide, where the density of molecules is relatively high. As our detection scheme relies on multi-photon transitions using a focused laser beam, a high density of molecules is crucial. In our configuration, the ions are extracted from the guide by applying a small positive voltage to two of the quadrupole rods. This results in a field that is homogeneous along the axis of the quadrupole while it defocuses the ions in the direction perpendicular to both the axis of the quadrupole and the axis of the ion optics. Here, we show how the astigmatism can be compensated with the use of a series of planar electrodes with horizontal and vertical slits.

\section{Experimental Setup}

\begin{figure*}[bth!]
\centering
  \includegraphics[width=17cm]{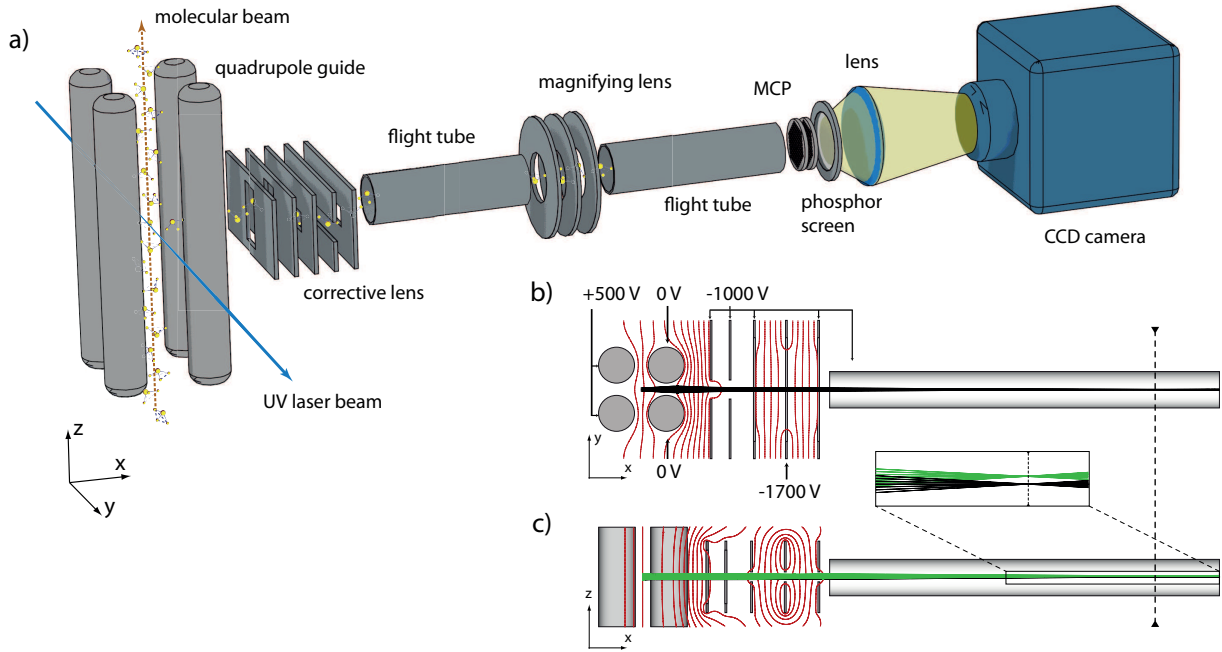}
  \caption{(a) Schematic view of the ion optics used for performing velocity map imaging inside a quadrupole guide. Rectangular electrodes with horizontal and vertical slits are used to make a position focus approximately 300\,mm from the center of the quadrupole. Using a cylindrical einzel lens this image is mapped onto a Multi Channel Plate detector mounted in front of a phosphor screen. The image is recorded using a CCD camera. (b) and (c) Equipotential lines in the lens system along the horizontal and vertical plane, respectively. The black and green lines show trajectories of ions starting with a velocity along the $z$-axis of 0 and 300\,m/s, respectively. }
  \label{fgr:setup}
\end{figure*}

Our experiments have been performed in a vertical molecular beam machine that is described in Ref~\citenum{Bethlem:EPJST2008}. 
In brief, a pulsed ($\sim$100\,$\mu$s) ammonia beam is released into vacuum from a solenoid valve (GeneralValve series 99) at a 10\,Hz repetition rate. By cooling the valve housing to -50$^{\circ}$\,C and seeding the ammonia molecules in xenon, the mean velocity of the beam is lowered to 300\,m/s. The ammonia beam is decelerated using a Stark decelerator consisting of an array of 101\,deceleration stages. Adjacent stages are 5.5\,mm apart. Each deceleration stage is formed by two parallel 3\,mm diameter cylindrical rods, spaced 2\,mm apart. The two opposite rods are switched to $+10$ and $-10$\,kV by four independent HV-switches that are triggered by a programmable delay generator. The velocity of the molecular beam is controlled completely by the computer generated burst sequence (for details on Stark deceleration, see Ref~\citenum{Bethlem:PRA2002,vandeMeerakker:ChemRev2012}). An electrostatic quadrupole, mounted 30\,mm behind the decelerator, is used for focusing the slow molecules and as extraction field for the velocity map imaging system. The chamber that houses the quadrupole guide is differentially pumped and kept at a pressure of $3\times10^{-8}$\,mbar when the pulsed valve is operating. 

A schematic view of the quadrupole and ion optics is shown in of Fig.~\ref{fgr:setup}(a). The quadrupole consists of four 20\,mm diameter cylindrical rods that are placed on the outside of a 20\,mm diameter circle. Alternating positive and negative voltages up to 6\,kV are applied to adjacent rods to focus decelerated ammonia molecules. The molecules are focused 90\,mm downstream from the decelerator, where they are ionized using a (2+1) resonance enhanced multi-photon ionization (REMPI) scheme. The laser radiation ($\sim$10~mJ in a 5~ns pulse @320\,nm) is focused between the rods of the quadrupole using a lens with a focal length of 500\,mm. In order to extract the ions, a positive voltage pulse is applied to the left hand side quadrupole rods while the other two rods are grounded. In this way, ions are pushed towards the corrective lens consisting of five rectangular (80$\times$40\,mm$^{2}$) electrodes, two of which having a vertical slit aperture (10$\times$30\,mm$^{2}$) and three having a horizontal slit aperture (50$\times$10\,mm$^{2}$). The fourth electrode is composed of two parts such that a voltage difference can be applied between the upper and lower part. The first electrode of the corrective lens is placed 40\,mm away from the center of the quadrupole, while the next electrodes are separated by 15\,mm. After a 220\,mm long flight tube, the ions pass a magnifying lens that consists of three circular electrodes with 15\,mm circular apertures \cite{Offerhaus:RevSciInstrum2001}. Finally, after another 240\,mm long flight tube, the ions impinge upon a double Multi Channel Plate (MCP) mounted in front of a fast response (1/e time is 10\,ns) Phosphor Screen (Photonis P-47 MgO). The light of the phosphor screen is imaged onto a CCD camera (PCO 1300, $1392\times1040$ pixels). By applying  a  timed  voltage  pulse  with  a  high-voltage switch (Behlke HTS 31-03-GSM), the gain of the front MCP can be gated to select ammonia ions, thus eliminating background signal due to oil and other residual molecules in the vacuum.

In Fig.~\ref{fgr:setup}(b-c), the equipotential lines in the lens system are shown along the horizontal and vertical plane, respectively. Also shown in the figure are the trajectories of ions, starting with a velocity along the $z$-axis of 0 (black lines) or 300\,m/s (green lines). These calculations were performed using the SIMION package\cite{Dahl:RevSciInstrum1990}. With voltages applied to the left hand side rods of the quadrupole, the ions are defocused in the horizontal plane (Fig.~\ref{fgr:setup}(b)). This is compensated by applying a voltage difference between the right hand side rods of the quadrupole and the first electrode. If an extraction voltage of +500\,V is applied, a voltage of -1000\,V is required on the first electrode to create a focus slightly before the magnifying lens. From our simulations, molecules starting with a velocity of 300\,m/s along the $z$-axis are displaced from the center by approximately 0.5\,mm. The magnifying lens is used to map the image plane onto the position sensitive detector, magnifying the image by a factor of about 6. In the vertical direction (Fig.~\ref{fgr:setup}(c)), the divergence of the beam is not affected by the quadrupole or the first two electrodes. The ions are focused onto the image plane by applying a voltage of -1700\,V to the fourth electrode while keeping the other electrodes and the flight tubes at a voltage of -1000\,V.

In our {\sc{vmi}} system, the focusing properties along the horizontal plane are determined by the voltage applied to the first electrode, while the focusing properties along the vertical plane are determined by the voltage applied to the fourth electrode. Note that the second and third electrodes are kept at the same voltage to ensure that the vertical and horizontal focusing properties are fully uncoupled. In addition, the position of the beam can be steered in the horizontal and vertical plane by applying small voltage differences (typically about 30\,V) between the two right hand side quadrupole rods and between the upper and lower part of the fourth electrode, respectively. 

\section{Characterizing the ion optics}

\begin{figure}[h!]
\centering
  \includegraphics[width=8.5cm]{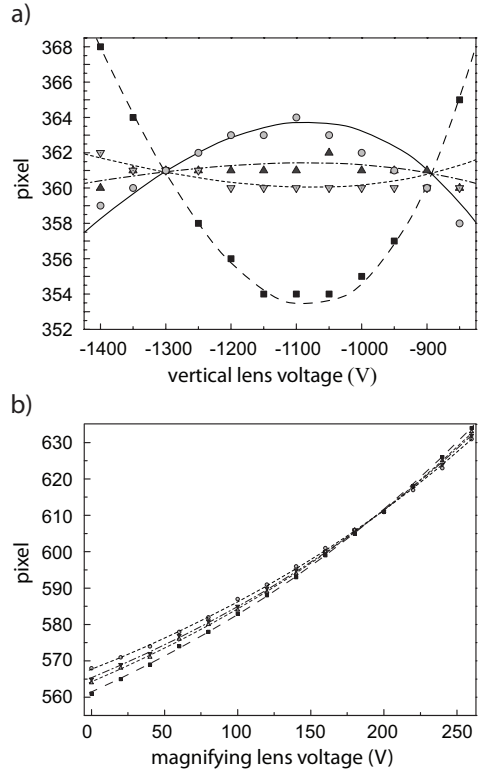}
  \caption{(a) Measured vertical position of the ions as a function of the voltage applied to the fourth electrode of the corrective lens with the laser focused at four different vertical positions in the quadrupole. A voltage of $-500$\,V is applied to the left hand side quadrupole rods while a voltage of $-1000$\,V is applied to all other electrodes. (b) Measured vertical position of the ions as a function of the voltage applied to the magnifying lens. The fourth electrode is kept at $-1700$\,V. The other voltages are set as in (a).}
  \label{fgr:focus}
\end{figure}

In order to have optimal velocity resolution, the voltages need to be applied such that ions with the same initial velocity but a different initial position are focused at the same position on the detector. In the $y$-direction, ions are collected over a range that is limited by the gap between the quadrupole rods. A position focus is found by changing the voltage applied to the left hand side rods until the observed ion distribution in the horizontal direction is minimized. To experimentally verify the optimal settings in the z-direction, we scanned the voltage applied to the fourth electrode while the laser focus was located at different heights. These measurements are shown as the symbols in Fig.~\ref{fgr:focus}. The curves in Fig.~\ref{fgr:focus} are the result of trajectory simulations using SIMION~\cite{Dahl:RevSciInstrum1990}. For the measurement shown in Fig.~\ref{fgr:focus}(a), the magnifying lens was not used. It is seen that, for an applied voltage around $-1100$\,V, the beam is neither focused nor defocused and the vertical position at which the ions arrive at the detector directly reflects the height of the laser focus (under our conditions, one pixel corresponds to about 0.05\,mm). Note that $-1100$\,V is slightly more negative than the expected $-1000$\,V that is applied to the other electrodes. This is attributed to the finite size of the electrodes. 

When the voltage applied to the fourth electrode is larger (more negative) than that of the surrounding electrodes, the ions are first accelerated and then decelerated as they pass the fourth electrode. At the same time, they experience a focusing, a defocusing and again a focusing force. As the ions are (on average) faster and closer to the axis when they experience a defocusing force and slower and further away from the axis when they experience a focusing force, the overall effect of the field is focusing.
This is the basic operation principle of an Einzel lens\cite{Whitaker:Book}. The same argument holds for the situation when the voltage applied to the fourth electrode is smaller (less negative) than that applied to the surrounding electrodes. As a result of this effect, position foci are observed at voltages of $-900$\,V and $-1300$\,V. Either voltage setting can be used to perform velocity map imaging; we chose the lower voltage setting. 

When the magnifying lens is used, the focal length of the first horizontal and vertical lens needs to be reduced. When a voltage of $+550$\,V is applied to the left hand side rods of the quadrupole and a voltage of $-1730$\,V is applied to the fourth electrode, the focal plane is situated about 30\,mm before the magnifying lens. It is seen from the measurements and simulations shown in Fig.~\ref{fgr:focus}(b) that the focal plane is imaged onto the detector when a voltage of $+200$\,V is applied to the middle electrode of the magnifying lens. 

\section{Experimental results}

\begin{figure}[h!]
\centering
  \includegraphics[width=8.5cm]{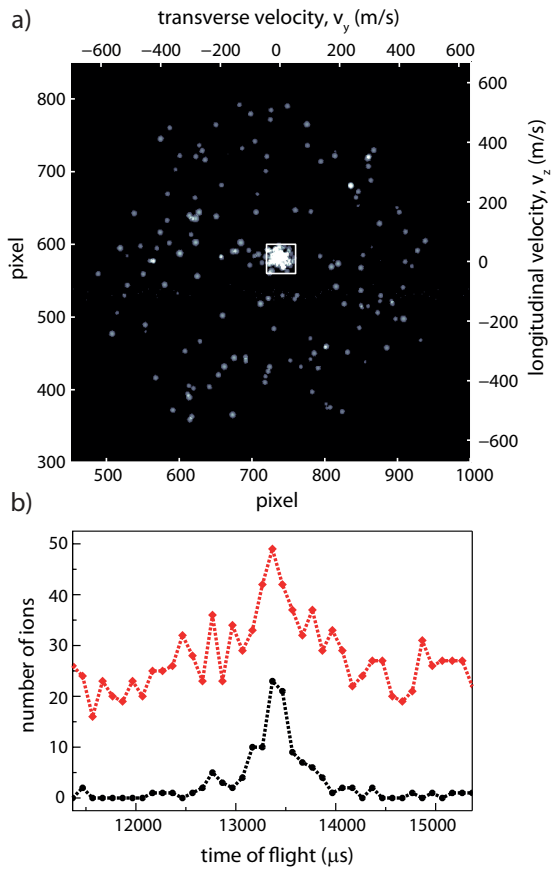}
  \caption{(a) Camera image showing the distribution of ions. The MCP gate is set to detect ammonia ions only. The bright spot inside the square corresponds to ammonia molecules that are decelerated from 300 to 10\,m/s. (b) Time-of-flight distributions for a molecular beam decelerated to 10\,m/s. The red curve is obtained by counting ions over the entire area of the camera, while the black curve is obtained by counting ions inside the white square depicted in (a).}
  \label{fgr:background_reduction}
\end{figure}

In Fig.~\ref{fgr:background_reduction}(a), a typical camera image is shown with the voltages set to obtain optimal velocity resolution. In this measurement, the decelerator is set to decelerate ammonia molecules from 300\,m/s to 10\,m/s. The image is averaged over 256 shots.
For each laser shot, pixels below a certain threshold are set to zero before adding up the images. The bright spot at the center of the image corresponds to the decelerated molecules, while the scattered spots result from thermal background ammonia molecules in the vacuum. Note that the ion distribution is clipped by the aperture of the magnifying lens. On the upper side and the right hand side, the horizontal and vertical position is translated into horizontal and vertical velocity, respectively (vide infra). 

In Fig.~\ref{fgr:background_reduction}(b), the observed ion signal is shown as a function of time with respect to the start of the burst sequence applied to the decelerator. The lower trace is obtained by counting the ions over the area enclosed by the $100 \times100\,$m$^{2}/$s$^{2}$ white square shown in Fig.~\ref{fgr:background_reduction}(a), while the upper trace is obtained by counting the ions over the entire area of the camera. Ions are counted using the centroiding method introduced in Ref~\citenum{Chang:RevSciInstrum1998}. Each data point is averaged over 32 shots. If ions are counted over the entire area of the camera, typically some 20 background ions are detected. By only counting ions within the selected area, this background is reduced by a factor of $550^{2}\pi/100^{2} \approx 100$. Note that for a given voltage the quadrupole will focus molecules within a small velocity interval. As a result, the width of the time of flight profile does not reflect the velocity spread of the decelerated packet.    

\begin{figure}[h!]
\centering
  \includegraphics[width=9cm]{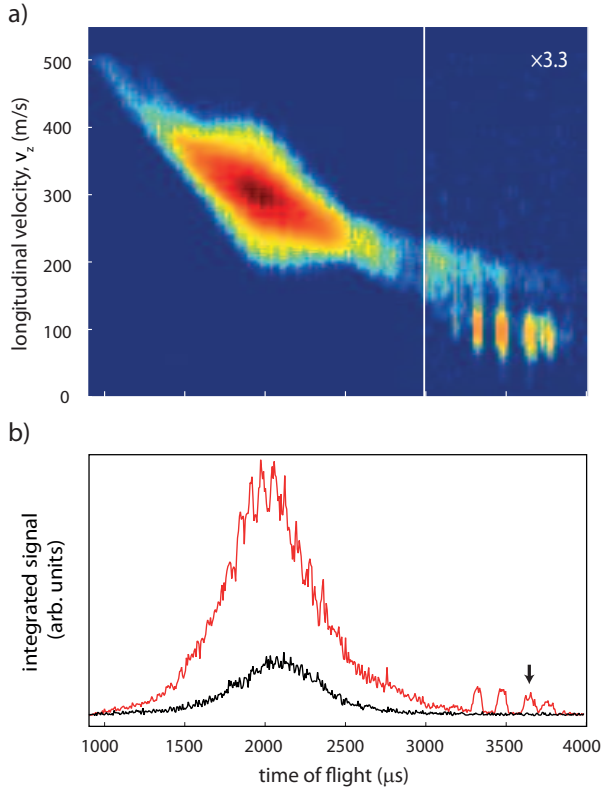}
  \caption{(a) Longitudinal velocity of the ammonia beam as a function of time. The intensity of the right hand side panel is scaled by a factor of 3.3 for clarity. (b) Integrated ion signal as a function of time with the decelerator on (red curve) and with the decelerator off (black curve). The arrow indicates the position of the synchronous packet of molecules.}
  \label{fgr:phasespace}
\end{figure}

In Fig.~\ref{fgr:phasespace}, the longitudinal velocity of the molecular beam is shown as a function of time with respect to the start of the burst sequence applied to the decelerator. For this image, the decelerator was set to decelerate from 300 to 100 m/s. At every time delay, an image is averaged over 64 shots and the intensity is integrated over the horizontal direction of the camera. This results in 156 vertical lines which are combined to obtain Fig.~\ref{fgr:phasespace}(a). The observed signal follows a hyperbolic curve due to the inverse relation between velocity and time. The most intense part of the beam is observed at short times having an average velocity of 300\,m/s.  These are molecules that were not in sync with the time sequence and, as a result, they are on average neither decelerated nor accelerated but they do experience a focusing effect in the transverse directions. At later arrival times, four packets are seen that are offset from this hyperbola. The packet arriving at 3700\,$\mu$s originates from molecules that were at the right position near the entrance of the decelerator at the start of the time sequence and that passed through all 101 deceleration stages in sync with the deceleration fields. These molecules entered the decelerator with a velocity of 300\,m/s and exit the decelerator with a velocity of 100\,m/s. The earlier peaks originate from molecules that were either one or two periods further inside the decelerator at the start of the time sequence. These molecules also entered the decelerator with a velocity of 300\,m/s, but since they missed the last two or four deceleration stages they exit the decelerator with a velocity of 108 or 115\,m/s, respectively. The peak that appears at a later time in the time-of-flight distribution originates from molecules entering the decelerator with a lower initial velocity, catching up with the time sequence one period later. Throughout the decelerator, it trails the synchronous packet by 11\,mm, exiting the decelerator with the same final velocity of 100\,m/s. Although the resolution of our {\sc{vmi}} system is insufficient to resolve the difference in velocity between the distinct packets, a small displacement can be observed.

In Fig.~\ref{fgr:phasespace}(b), the ion signal is integrated over the total area of the camera. The red curve shows the time-of-flight with the decelerator on, while the black curve shows the same measurement taken with the decelerator off. Note that the observed oscillations in the undecelerated part of the beam are due to velocity modulations, and are well reproduced in simulations\cite{vandeMeerakker:ChemRev2012}.

\begin{figure}[h!]
\centering
  \includegraphics[width=9cm]{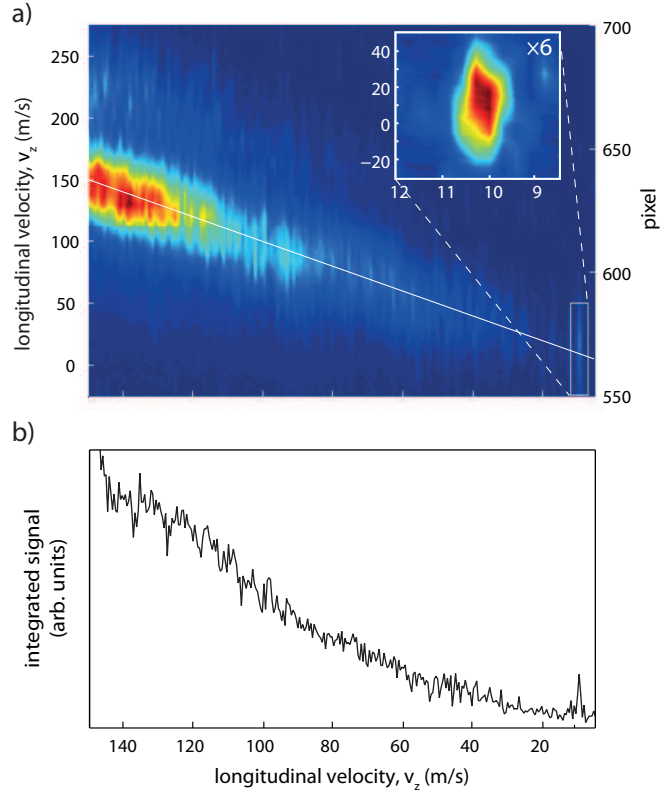}
  \caption{(a) Longitudinal velocity of the ammonia beam as a function of the velocity set by the burst sequence applied to the decelerator. The quadrupole is set to focus molecules of 10\,m/s at the position of the laser. The inset shows an enlarged 
view of molecules that are focused with the quadrupole. The intensity is scaled by a factor of 6 for clarity. (b) Integrated ion signal as a function of velocity with the burst sequence set to decelerate molecules to different velocities.}
  \label{fgr:velocityscan}
\end{figure}

In order to calibrate the velocity axis, we have recorded {\sc{vmi}} images while the velocity of the beam was scanned from 150 to 5\,m/s in steps of 1\,m/s by changing the burst sequence applied to the Stark decelerator. At every velocity, an image is averaged over 64 shots and the intensity is integrated over the horizontal direction of the camera. This results in 145 vertical lines which are combined to obtain Fig.~\ref{fgr:velocityscan}(a). Since molecules are ionized using a focused laser beam, the number of detected ions is proportional to the density of the decelerated beam. Due to the increasing time of flight, this density decreases rapidly when the velocity of the beam is lowered. In these measurements, the quadrupole is set to transversally focus molecules of 10\,m/s at the position of the laser. As a result, a bright spot at this velocity is observed in Fig.~\ref{fgr:velocityscan}(a) (shown enlarged in the inset). The straight line also shown in the figure is used to translate the position at the camera into a vertical velocity. The FWHM velocity spread of the beam that is inferred from the camera images is 35\,m/s (corresponding to 20 pixels on the camera). As, at the used settings of the decelerator, the longitudinal velocity spread of the beam is below 10\,m/s~\cite{Bethlem:PRA2002}, we conclude that the velocity resolution of the lens system is on the order of 35\,m/s. 

In Fig.~\ref{fgr:velocityscan}(b), the ion signal is integrated over the total area of the camera. A small peak is observed when molecules are decelerated to 10\,m/s corresponding to molecules that are focused 90\,mm downstream from the decelerator, where the molecular beam crosses the ionisation laser. 

\section{Conclusions}

A setup consisting of a series of planar electrodes is used for performing velocity map imaging inside an electrostatic quadrupole guide. We obtain a velocity resolution of 35\,m/s, limited by the size of the electrodes and by the presence of isolators used for suspending the quadrupole. Our work demonstrates that velocity map imaging can be performed in unconventional electrode configurations 
while achieving a resolution that is sufficient for most purposes. This should be useful for studies in electrostatic traps\cite{Bethlem:PRA2002}, storage rings\cite{Heiner:NatPhys2007,Zieger:PRL2010} and molecular fountains\cite{Bethlem:EPJST2008}.
An often quoted phrase in the ion imaging community says: "if you do not use {\sc{vmi}}, you throw away information". 
In this work, the information obtained from {\sc{vmi}} is used to detect slow molecules free of any signal from thermal background gas.

\section*{Acknowledgements}

This research has been supported by NWO via a VIDI-grant and by the ERC via a Starting Grant. We acknowledge the expert technical assistance of Jacques Bouma. We thank Wim Ubachs for continuing interest and support.

\balance


\footnotesize{
\providecommand*{\mcitethebibliography}{\thebibliography}
\csname @ifundefined\endcsname{endmcitethebibliography}
{\let\endmcitethebibliography\endthebibliography}{}

}
\begin{mcitethebibliography}{17}
\providecommand*{\natexlab}[1]{#1}
\providecommand*{\mciteSetBstSublistMode}[1]{}
\providecommand*{\mciteSetBstMaxWidthForm}[2]{}
\providecommand*{\mciteBstWouldAddEndPuncttrue}
  {\def\EndOfBibitem{\unskip.}}
\providecommand*{\mciteBstWouldAddEndPunctfalse}
  {\let\EndOfBibitem\relax}
\providecommand*{\mciteSetBstMidEndSepPunct}[3]{}
\providecommand*{\mciteSetBstSublistLabelBeginEnd}[3]{}
\providecommand*{\EndOfBibitem}{}
\mciteSetBstSublistMode{f}
\mciteSetBstMaxWidthForm{subitem}
{(\emph{\alph{mcitesubitemcount}})}
\mciteSetBstSublistLabelBeginEnd{\mcitemaxwidthsubitemform\space}
{\relax}{\relax}

\bibitem[Hudson \emph{et~al.}(2011)Hudson, Kara, Smallman, Sauer, Tarbutt, and
  Hinds]{Hudson:Nature2011}
J.~J. Hudson, D.~M. Kara, I.~J. Smallman, B.~E. Sauer, M.~R. Tarbutt and E.~A.
  Hinds, \emph{Nature}, 2011, \textbf{473}, 493--6\relax
\mciteBstWouldAddEndPuncttrue
\mciteSetBstMidEndSepPunct{\mcitedefaultmidpunct}
{\mcitedefaultendpunct}{\mcitedefaultseppunct}\relax
\EndOfBibitem
\bibitem[Tarbutt \emph{et~al.}(2009)Tarbutt, Hudson, Sauer, and
  Hinds]{Tarbutt:FarDis2009}
M.~R. Tarbutt, J.~J. Hudson, B.~E. Sauer and E.~A. Hinds, \emph{Faraday
  Discussions}, 2009, \textbf{142}, 37\relax
\mciteBstWouldAddEndPuncttrue
\mciteSetBstMidEndSepPunct{\mcitedefaultmidpunct}
{\mcitedefaultendpunct}{\mcitedefaultseppunct}\relax
\EndOfBibitem
\bibitem[DeMille \emph{et~al.}(2008)DeMille, Cahn, Murphree, Rahmlow, and
  Kozlov]{DeMille:PRL2008}
D.~DeMille, S.~B. Cahn, D.~Murphree, D.~A. Rahmlow and M.~G. Kozlov,
  \emph{Physical Review Letters}, 2008, \textbf{100}, 1--4\relax
\mciteBstWouldAddEndPuncttrue
\mciteSetBstMidEndSepPunct{\mcitedefaultmidpunct}
{\mcitedefaultendpunct}{\mcitedefaultseppunct}\relax
\EndOfBibitem
\bibitem[Bethlem \emph{et~al.}(2008)Bethlem, Kajita, Sartakov, Meijer, and
  Ubachs]{Bethlem:EPJST2008}
H.~L. Bethlem, M.~Kajita, B.~Sartakov, G.~Meijer and W.~Ubachs, \emph{The
  European Physical Journal Special Topics}, 2008, \textbf{163}, 55--69\relax
\mciteBstWouldAddEndPuncttrue
\mciteSetBstMidEndSepPunct{\mcitedefaultmidpunct}
{\mcitedefaultendpunct}{\mcitedefaultseppunct}\relax
\EndOfBibitem
\bibitem[van Veldhoven \emph{et~al.}(2004)van Veldhoven, K\"{u}pper, Bethlem,
  Sartakov, Roij, and Meijer]{vanVeldhoven:EPJD2004}
J.~van Veldhoven, J.~K\"{u}pper, H.~L. Bethlem, B.~Sartakov, A.~J.~A. Roij and
  G.~Meijer, \emph{The European Physical Journal D}, 2004, \textbf{31},
  337--349\relax
\mciteBstWouldAddEndPuncttrue
\mciteSetBstMidEndSepPunct{\mcitedefaultmidpunct}
{\mcitedefaultendpunct}{\mcitedefaultseppunct}\relax
\EndOfBibitem
\bibitem[Hudson \emph{et~al.}(2006)Hudson, Lewandowski, Sawyer, and
  Ye]{Hudson:PRL2006}
E.~R. Hudson, H.~J. Lewandowski, B.~C. Sawyer and J.~Ye, \emph{Physical Review
  Letters}, 2006, \textbf{96}, 1--4\relax
\mciteBstWouldAddEndPuncttrue
\mciteSetBstMidEndSepPunct{\mcitedefaultmidpunct}
{\mcitedefaultendpunct}{\mcitedefaultseppunct}\relax
\EndOfBibitem
\bibitem[Eppink and Parker(1997)]{Eppink:RevSciInstrum1997}
A.~T. J.~B. Eppink and D.~H. Parker, \emph{Review of Scientific Instruments},
  1997, \textbf{68}, 3477\relax
\mciteBstWouldAddEndPuncttrue
\mciteSetBstMidEndSepPunct{\mcitedefaultmidpunct}
{\mcitedefaultendpunct}{\mcitedefaultseppunct}\relax
\EndOfBibitem
\bibitem[Chandler and Houston(1987)]{Chandler:JCP1987}
D.~W. Chandler and P.~L. Houston, \emph{The Journal of Chemical Physics}, 1987,
  \textbf{87}, 1445\relax
\mciteBstWouldAddEndPuncttrue
\mciteSetBstMidEndSepPunct{\mcitedefaultmidpunct}
{\mcitedefaultendpunct}{\mcitedefaultseppunct}\relax
\EndOfBibitem
\bibitem[{B. J. Whitaker}(2003)]{Whitaker:Book}
{B. J. Whitaker}, \emph{{Imaging in Molecular Dynamics: Technology and
  Applications}}, Cambridge University Press, Cambridge, 2003\relax
\mciteBstWouldAddEndPuncttrue
\mciteSetBstMidEndSepPunct{\mcitedefaultmidpunct}
{\mcitedefaultendpunct}{\mcitedefaultseppunct}\relax
\EndOfBibitem
\bibitem[Lipciuc \emph{et~al.}(2006)Lipciuc, Buijs, and
  Janssen]{Lipciuc:PCCP2006}
M.~L. Lipciuc, J.~B. Buijs and M.~H.~M. Janssen, \emph{Physical Chemistry
  Chemical Physics}, 2006, \textbf{8}, 219--23\relax
\mciteBstWouldAddEndPuncttrue
\mciteSetBstMidEndSepPunct{\mcitedefaultmidpunct}
{\mcitedefaultendpunct}{\mcitedefaultseppunct}\relax
\EndOfBibitem
\bibitem[Bethlem \emph{et~al.}(2002)Bethlem, Crompvoets, Jongma, van~de
  Meerakker, and Meijer]{Bethlem:PRA2002}
H.~L. Bethlem, F.~M.~H. Crompvoets, R.~T. Jongma, S.~Y.~T. van~de Meerakker and
  G.~Meijer, \emph{Physical Review A}, 2002, \textbf{65}, 1--20\relax
\mciteBstWouldAddEndPuncttrue
\mciteSetBstMidEndSepPunct{\mcitedefaultmidpunct}
{\mcitedefaultendpunct}{\mcitedefaultseppunct}\relax
\EndOfBibitem
\bibitem[van~de Meerakker \emph{et~al.}(2012)van~de Meerakker, Bethlem,
  Vanhaecke, and Meijer]{vandeMeerakker:ChemRev2012}
S.~Y.~T. van~de Meerakker, H.~L. Bethlem, N.~Vanhaecke and G.~Meijer,
  \emph{Chemical Reviews}, 2012\relax
\mciteBstWouldAddEndPuncttrue
\mciteSetBstMidEndSepPunct{\mcitedefaultmidpunct}
{\mcitedefaultendpunct}{\mcitedefaultseppunct}\relax
\EndOfBibitem
\bibitem[Offerhaus \emph{et~al.}(2001)Offerhaus, Nicole, Lépine, Bordas,
  Rosca-Pruna, and Vrakking]{Offerhaus:RevSciInstrum2001}
H.~L. Offerhaus, C.~Nicole, F.~Lépine, C.~Bordas, F.~Rosca-Pruna and M.~J.~J.
  Vrakking, \emph{Review of Scientific Instruments}, 2001, \textbf{72},
  3245\relax
\mciteBstWouldAddEndPuncttrue
\mciteSetBstMidEndSepPunct{\mcitedefaultmidpunct}
{\mcitedefaultendpunct}{\mcitedefaultseppunct}\relax
\EndOfBibitem
\bibitem[Dahl \emph{et~al.}(1990)Dahl, Delmore, and
  Appelhans]{Dahl:RevSciInstrum1990}
D.~A. Dahl, J.~E. Delmore and A.~D. Appelhans, \emph{Review of Scientific
  Instruments}, 1990, \textbf{61}, 607\relax
\mciteBstWouldAddEndPuncttrue
\mciteSetBstMidEndSepPunct{\mcitedefaultmidpunct}
{\mcitedefaultendpunct}{\mcitedefaultseppunct}\relax
\EndOfBibitem
\bibitem[Chang \emph{et~al.}(1998)Chang, Hoetzlein, Mueller, Geiser, and
  Houston]{Chang:RevSciInstrum1998}
B.~Y. Chang, R.~C. Hoetzlein, J.~A. Mueller, J.~D. Geiser and P.~L. Houston,
  \emph{Review of Scientific Instruments}, 1998, \textbf{69}, 1665--1670\relax
\mciteBstWouldAddEndPuncttrue
\mciteSetBstMidEndSepPunct{\mcitedefaultmidpunct}
{\mcitedefaultendpunct}{\mcitedefaultseppunct}\relax
\EndOfBibitem
\bibitem[Heiner \emph{et~al.}(2007)Heiner, Carty, Meijer, and
  Bethlem]{Heiner:NatPhys2007}
C.~E. Heiner, D.~Carty, G.~Meijer and H.~L. Bethlem, \emph{Nature Physics},
  2007, \textbf{3}, 115--118\relax
\mciteBstWouldAddEndPuncttrue
\mciteSetBstMidEndSepPunct{\mcitedefaultmidpunct}
{\mcitedefaultendpunct}{\mcitedefaultseppunct}\relax
\EndOfBibitem
\bibitem[Zieger \emph{et~al.}(2010)Zieger, van~de Meerakker, Heiner, Bethlem,
  van Roij, and Meijer]{Zieger:PRL2010}
P.~C. Zieger, S.~Y.~T. van~de Meerakker, C.~E. Heiner, H.~L. Bethlem, A.~J.~A.
  van Roij and G.~Meijer, \emph{Physical Review Letters}, 2010, \textbf{105},
  1--4\relax
\mciteBstWouldAddEndPuncttrue
\mciteSetBstMidEndSepPunct{\mcitedefaultmidpunct}
{\mcitedefaultendpunct}{\mcitedefaultseppunct}\relax
\EndOfBibitem
\end{mcitethebibliography}
\end{document}